\documentstyle[12pt,epsfig]{article}
 
\catcode`\@=11
\long\def\@makefntext#1{ 
\protect\noindent \hbox to 3.2pt {\hskip-.9pt
$^{{\ninerm\@thefnmark}}$\hfil}#1\hfill} 
 
\def\thefootnote{\fnsymbol{footnote}}
 \def\@makefnmark{\hbox to 0pt{$^{\@thefnmark}$\hss}}  
 
\def\ps@myheadings{\let\@mkboth\@gobbletwo
\def\@oddhead{\hbox{} 
\rightmark\hfil\ninerm\thepage}
\def\@oddfoot{}\def\@evenhead{\ninerm\thepage\hfil 
\leftmark\hbox{}}\def\@evenfoot{}
\def\sectionmark##1{}\def\subsectionmark##1{}}
 
\newcommand{\elevenbf}{\bf}
\newcommand{\elevenit}{\it}
\newcommand{\beq}{\begin{equation}}
\newcommand{\eeq}{\end{equation}}
\newcommand{\bea}{\begin{eqnarray}}
\newcommand{\eea}{\end{eqnarray}}
\newcommand{\barr}{\begin{array}}
\newcommand{\earr}{\end{array}}
\newcommand{\bc}{\begin{center}}
\newcommand{\ec}{\end{center}}
\newcommand{\btab}{\begin{tabular}}
\newcommand{\etab}{\end{tabular}}
\newcommand{\gv}{\mbox{GeV}}

\newcommand{\nn}{\nonumber}

\newcommand{\al}{\alpha}

\newcommand{\G}{\Gamma}
\newcommand{\Gmu}{G_{\mu}}

\newcommand{\ganu}{\gamma_{\nu}}

\newcommand{\gafi}{\gamma_5}

\newcommand{\noi}{\noindent}

\newcommand{\epm}{e^+e^-}

\newcommand{\sm}{Standard Model }

\newcommand{\mz}{M_Z^2}

\newcommand{\Dr}{\Delta r}

\newcommand{\alr}{A_{LR}}

\newcommand{\ass}{asymmetries }

\newcommand{\zp}{{\it Z.\ Phys.\ }}

\newcommand{\prl}{{\it Phys.\ Rev.\ Lett.\ }}
\newcommand{\np}{{\it Nucl.\ Phys.\ }}
\newcommand{\plb}{{\it Phys.\ Lett.\ }}
\newcommand{\prd}{{\it Phys.\ Rev.\  }}

\begin{document}
 
\newcommand{\symbolfootnote}{\renewcommand{\thefootnote}
        {\fnsymbol{footnote}}}
\renewcommand{\thefootnote}{\fnsymbol{footnote}}
\newcommand{\alphfootnote}
        {\setcounter{footnote}{0}
         \renewcommand{\thefootnote}{\sevenrm\alph{footnote}}}
 
\newcounter{sectionc}\newcounter{subsectionc}\newcounter{subsubsectionc}
\renewcommand{\section}[1] {\vspace{0.6cm}\addtocounter{sectionc}{1}
\setcounter{subsectionc}{0}\setcounter{subsubsectionc}{0}\noindent
        {\bf\thesectionc. #1}\par\vspace{0.4cm}}
\renewcommand{\subsection}[1] {\vspace{0.6cm}\addtocounter{subsectionc}{1}
        \setcounter{subsubsectionc}{0}\noindent
        {\it\thesectionc.\thesubsectionc. #1}\par\vspace{0.4cm}}
\renewcommand{\subsubsection}[1] {\vspace{0.6cm}\addtocounter{subsubsectionc}{1}
        \noindent {\rm\thesectionc.\thesubsectionc.\thesubsubsectionc.
        #1}\par\vspace{0.4cm}}
\newcommand{\nonumsection}[1] {\vspace{0.6cm}\noindent{\bf #1}
        \par\vspace{0.4cm}}
 
\newcounter{appendixc}
\newcounter{subappendixc}[appendixc]
\newcounter{subsubappendixc}[subappendixc]
\renewcommand{\thesubappendixc}{\Alph{appendixc}.\arabic{subappendixc}}
\renewcommand{\thesubsubappendixc}
        {\Alph{appendixc}.\arabic{subappendixc}.\arabic{subsubappendixc}}
 
\renewcommand{\appendix}[1] {\vspace{0.6cm}
        \refstepcounter{appendixc}
        \setcounter{figure}{0}
        \setcounter{table}{0}
        \setcounter{equation}{0}
        \renewcommand{\thefigure}{\Alph{appendixc}.\arabic{figure}}
        \renewcommand{\thetable}{\Alph{appendixc}.\arabic{table}}
        \renewcommand{\theappendixc}{\Alph{appendixc}}
        \renewcommand{\theequation}{\Alph{appendixc}.\arabic{equation}}
        \noindent{\bf Appendix \theappendixc #1}\par\vspace{0.4cm}}
\newcommand{\subappendix}[1] {\vspace{0.6cm}
        \refstepcounter{subappendixc}
        \noindent{\bf Appendix \thesubappendixc. #1}\par\vspace{0.4cm}}
\newcommand{\subsubappendix}[1] {\vspace{0.6cm}
        \refstepcounter{subsubappendixc}
        \noindent{\it Appendix \thesubsubappendixc. #1}
        \par\vspace{0.4cm}}
 
\def\abstracts#1{{
        \centering{\begin{minipage}{30pc}\tenrm\baselineskip=12pt\noindent
        \centerline{\tenrm ABSTRACT}\vspace{0.3cm}
        \parindent=0pt #1
        \end{minipage} }\par}}
 
\newcommand{\bibit}{\it}
\newcommand{\bibbf}{\bf}
\renewenvironment{thebibliography}[1]
        {\begin{list}{\arabic{enumi}.}
        {\usecounter{enumi}\setlength{\parsep}{0pt}
\setlength{\leftmargin 1.25cm}{\rightmargin 0pt}
         \setlength{\itemsep}{0pt} \settowidth
        {\labelwidth}{#1.}\sloppy}}{\end{list}}
 
\topsep=0in\parsep=0in\itemsep=0in
\parindent=1.5pc
 
\newcounter{itemlistc}
\newcounter{romanlistc}
\newcounter{alphlistc}
\newcounter{arabiclistc}
\newenvironment{itemlist}
        {\setcounter{itemlistc}{0}
         \begin{list}{$\bullet$}
        {\usecounter{itemlistc}
         \setlength{\parsep}{0pt}
         \setlength{\itemsep}{0pt}}}{\end{list}}
 
\newenvironment{romanlist}
        {\setcounter{romanlistc}{0}
         \begin{list}{$($\roman{romanlistc}$)$}
        {\usecounter{romanlistc}
         \setlength{\parsep}{0pt}
         \setlength{\itemsep}{0pt}}}{\end{list}}
 
\newenvironment{alphlist}
        {\setcounter{alphlistc}{0}
         \begin{list}{$($\alph{alphlistc}$)$}
        {\usecounter{alphlistc}
         \setlength{\parsep}{0pt}
         \setlength{\itemsep}{0pt}}}{\end{list}}
 
\newenvironment{arabiclist}
        {\setcounter{arabiclistc}{0}
         \begin{list}{\arabic{arabiclistc}}
        {\usecounter{arabiclistc}
         \setlength{\parsep}{0pt}
         \setlength{\itemsep}{0pt}}}{\end{list}}
 
\newcommand{\fcaption}[1]{
        \refstepcounter{figure}
        \setbox\@tempboxa = \hbox{\tenrm Fig.~\thefigure. #1}
        \ifdim \wd\@tempboxa > 6in
           {\begin{center}
        \parbox{6in}{\tenrm\baselineskip=12pt Fig.~\thefigure. #1 }
            \end{center}}
        \else
             {\begin{center}
             {\tenrm Fig.~\thefigure. #1}
              \end{center}}
        \fi}
 
\newcommand{\tcaption}[1]{
        \refstepcounter{table}
        \setbox\@tempboxa = \hbox{\tenrm Table~\thetable. #1}
        \ifdim \wd\@tempboxa > 6in
           {\begin{center}
        \parbox{6in}{\tenrm\baselineskip=12pt Table~\thetable. #1 }
            \end{center}}
        \else
             {\begin{center}
             {\tenrm Table~\thetable. #1}
              \end{center}}
        \fi}
 
\def\@citex[#1]#2{\if@filesw\immediate\write\@auxout
        {\string\citation{#2}}\fi
\def\@citea{}\@cite{\@for\@citeb:=#2\do
        {\@citea\def\@citea{,}\@ifundefined
        {b@\@citeb}{{\bf ?}\@warning
        {Citation `\@citeb' on page \thepage \space undefined}}
        {\csname b@\@citeb\endcsname}}}{#1}}
 
\newif\if@cghi
\def\cite{\@cghitrue\@ifnextchar [{\@tempswatrue
        \@citex}{\@tempswafalse\@citex[]}}
\def\citelow{\@cghifalse\@ifnextchar [{\@tempswatrue
        \@citex}{\@tempswafalse\@citex[]}}
\def\@cite#1#2{{$\null^{#1}$\if@tempswa\typeout
        {IJCGA warning: optional citation argument
        ignored: `#2'} \fi}}
\newcommand{\citeup}{\cite}
 
\def\fnm#1{$^{\mbox{\scriptsize #1}}$}
\def\fnt#1#2{\footnotetext{\kern-.3em
        {$^{\mbox{\sevenrm #1}}$}{#2}}}
 
\font\twelvebf=cmbx10 scaled\magstep 1
\font\twelverm=cmr10 scaled\magstep 1
\font\twelveit=cmti10 scaled\magstep 1
\font\elevenbfit=cmbxti10 scaled\magstephalf
\font\elevenbf=cmbx10 scaled\magstephalf
\font\elevenrm=cmr10 scaled\magstephalf
\font\elevenit=cmti10 scaled\magstephalf
\font\bfit=cmbxti10
\font\tenbf=cmbx10
\font\tenrm=cmr10
\font\tenit=cmti10
\font\ninebf=cmbx9
\font\ninerm=cmr9
\font\nineit=cmti9
\font\eightbf=cmbx8
\font\eightrm=cmr8
\font\eightit=cmti8
 
 
\centerline{\tenbf SIGNALS FROM SUPERSYMMETRY IN  
                   ELECTROWEAK PRECISION DATA? } 
\vspace{0.8cm}
\centerline{\tenrm WOLFGANG HOLLIK }
\baselineskip=13pt
\centerline{\tenit Institut f\"ur Theoretische Physik,
                   Universit\"at Karlsruhe}
\baselineskip=12pt
\centerline{\tenit D-76128 Karlsruhe, Germany} \hfill \\[0.5cm]
\abstracts{The predictions of the \sm and the minimal supersymmetric
standard model (MSSM) for the electroweak precision parameters are discussed
in the light of the recent precision data. The results from global fits 
yield lower $\chi^2$ values in the MSSM than in the Standard model.
The fits prefer regions in the MSSM parameter space with
 $M_2 \simeq |\mu| $ and allow chargino masses higher than the present
exclusion limits of LEP 1.5.}

\vspace*{0.5cm}
\twelverm   
\baselineskip=14pt

\section{Introduction}
The present generation of high  precision experiments imposes
stringent tests on the Standard Model
 and its possible extensions.
Besides the impressive achievements in the determination of the
$Z$ boson parameters \cite{lep} and the $W$ mass
\cite{wmass}, the most important step has been the confirmation
of the top quark at the Tevatron \cite{top} with the mass
average value
$            m_t = 180 \pm 12$ GeV.

\smallskip
 The lack
of direct signals from ``New Physics'' makes the high
precision experiments
 also a unique tool in the search for {\it indirect} effects:
through  deviations of the experimental results
from the theoretical predictions of the minimal Standard Model.
 We discuss  
 the minimal supersymmetric standard model
 as a special example of
particular theoretical interest.

\section{Electroweak precision observables}
The $Z$ boson observables and the $W$ mass are conveniently calculated
 in terms of effective couplings at the $Z$ peak and the 
quantity $\Dr$ in the correlation between $M_{W,Z}$ and the Fermi
constant $\Gmu$. The formal relations are identical for the Standard
Model and the MSSM. For details on the \sm calculations see
\cite{yb95}, and for calculations in the MSSM we refer to ref.~\cite{susy5}.

\bigskip
\noindent
{\it Effective $Z$ boson couplings:}
The effective couplings follow
from the set of 1-loop diagrams
without virtual photons,
the non-QED  or weak  corrections.
These weak corrections
can conveniently be written
in terms of fermion-dependent overall normalizations
$\rho_f$ and effective mixing angles $s_f^2$
in the NC vertices, which contain the details of the models:
\bea
 & &
 J_{\nu}^{NC}  =   \left( \sqrt{2}\Gmu\mz \right)^{1/2} \,
  (g_V^f \,\ganu -  g_A^f \,\ganu\gafi)  \\
 & &
   =  \left( \sqrt{2}\Gmu\mz \rho_f \right)^{1/2}
\left( (I_3^f-2Q_fs_f^2)\ganu-I_3^f\ganu\gafi \right)  . \nn
\eea
 
\smallskip
\paragraph{\it Asymmetries and mixing angles:}
 
The effective mixing angles are of particular interest since
they determine the on-resonance asymmetries via the combinations
   \beq
    A_f = \frac{2g_V^f g_A^f}{(g_V^f)^2+(g_A^f)^2}  \, .
\eeq
Measurements of the \ass hence are measurements of
the ratios
\beq
  g_V^f/g_A^f = 1 - 2 Q_f s_f^2
\eeq
or the effective mixing angles, respectively.

\smallskip
\paragraph{\it $Z$ widths:}
  The fermionic partial
widths of the $Z$ boson,
 when
expressed in terms of the effective coupling constants
read up to 2nd order in the (light) fermion masses:
\bea
\Gamma_f
  & = & \G_0
 \, \left(
     (g_V^f)^2  +
     (g_A^f)^2 (1-\frac{6m_f^2}{\mz} )
                           \right)
 \cdot   (1+ Q_f^2\, \frac{3\al}{4\pi} ) \nn \\
     & +& \Delta\G^f_{QCD} \nn
\eea
with
$$
\G_0 \, =\,
  N_C^f\,\frac{\sqrt{2}\Gmu M_Z^3}{12\pi},
 \;\;\;\; N_C^f = 1
 \mbox{ (leptons)}, \;\; = 3 \mbox{ (quarks)}.
$$
 
\paragraph{\it The $W$ mass:}
The correlation between
the masses $M_W,M_Z$ of the vector bosons          in terms
of the Fermi constant $\Gmu$ is given by:
\beq
\frac{\Gmu}{\sqrt{2}}   =
            \frac{\pi\al}{2s_W^2 M_W^2} \frac{1}{1-\Dr}
\eeq
with the higher order quantity $\Dr$, containing the details of the 
 models.
 
\section{Standard model predictions versus data}
In table 1
the \sm predictions for $Z$ pole observables and the $W$ mass  are
put together. The first error corresponds to
the variation of $m_t$ in the observed range (1) and $ 60 < M_H < 1000$ GeV.
The second error is the hadronic
uncertainty from $\al_s=0.123\pm 0.006$, as measured
by QCD observables at the $Z$ \cite{alfas}.
 The recent combined LEP results \cite{lep} on the $Z$ resonance
parameters, under the assumption of lepton universality,
are also shown in table 1, together with $s^2_e$ from
the left-right asymmetry at the SLC \cite{sld}.

\begin{table*}[t]
            \caption{Precision observables: experimental results
             (from refs.\ 1,2,3)
             and standard model         
             predictions. } \vspace{0.5cm}
            \bc
 \btab{|| l | l | r || }
\hline
\hline
 observable & exp. (1995) & \sm prediction \\
\hline
\hline
$M_Z$ (GeV) & $91.1884\pm0.0022$ &  input \\
\hline
$\Gamma_Z$ (GeV) & $2.4963\pm 0.0032$ & $2.4976 \pm 0.0077\pm 0.0033$ \\
\hline
$\sigma_0^{had}$ (nb) & $41.4882\pm 0.078$ & $41.457\pm0.011\pm0.076$ \\
\hline
 $\G_{had}/\G_e$ & $20.788\pm 0.032 $ & $20.771\pm 0.019\pm 0.038$ \\
\hline
$\Gamma_{inv}$ (MeV) & $499.9\pm 2.5$ & $501.6\pm 1.1$ \\
\hline
$\G_b/\G_{had}=R_b$  & $0.2219\pm 0.0017$ & $0.2155\pm 0.0004$ \\
\hline
$\G_c/\G_{had}=R_c$  & $0.1540\pm 0.0074$ & $0.1723\pm 0.0002$ \\
  \hline
$A_b$            & $0.841\pm 0.053$  & $0.9346 \pm 0.0006$ \\
\hline
$\rho_{\ell}$ & $1.0044\pm 0.0016$ & $1.0050\pm 0.0023$ \\
\hline
$s^2_{\ell}$ (LEP) & $0.23186\pm 0.00034$ & $0.2317\pm 0.0012$ \\
\hline
$s^2_e (A_{LR})$ & $0.23049\pm 0.00050$ & $0.2317\pm 0.0012$   \\
 LEP$+$SLC   &  $0.23143\pm 0.00028$    &                    \\
\hline
$M_W$ (GeV) & $80.26 \pm 0.16$ & $80.36\pm 0.18$  \\
\hline
\hline
\etab
\ec 
\clearpage
\end{table*}

\smallskip \noindent
Significant deviations from the \sm predictions are observed in the
ratios  $R_b = \Gamma_b/\Gamma_{had}$ and
 $R_c = \Gamma_c/\Gamma_{had}$. The experimental values,
together with the top mass (1) from the Tevatron, are compatible
with the \sm at a confidence level of less than 1\% (see
\cite{lep}),
enough to claim a deviation from the Standard Model.
The other precision observables are in perfect agreement with the
Standard Model. Note that
the experimental value for $\rho_{\ell}$ exhibits the presence of
genuine electroweak corrections by nearly 3 standard deviations.
 
\smallskip
Assuming the validity of the \sm a global fit to all electroweak
results from LEP, SLD, $p\bar{p}$ and $\nu N$
constrains the parameters $m_t,\al_s$ as follows:
\cite{lep}:
\beq
    m_t = 178\pm 8^{+17}_{-20}\, \gv, \;\;\     
    \al_s = 0.123 \pm 0.004 \pm 0.002
\eeq
with $M_H= 300$ GeV for the central value.
The second error is from the variation of $M_H$
between 60 GeV and 1 TeV.
The fit results include the
uncertainties of the \sm calculations.

\section{The minimal supersymmetric standard model (MSSM):}
The MSSM deserves a special discussion
as the most predictive framework beyond the minimal model.
Its structure allows a similarly complete calculation of
the electroweak precision observables
as in the standard model in terms of one Higgs mass
(usually taken as $M_A$) and $\tan\beta= v_2/v_1$,
together with the set of
SUSY soft breaking parameters fixing the chargino/neutralino and
scalar fermion sectors.
It has been known since quite some time
\cite{higgs}
that light non-standard
Higgs bosons as well as light stop and charginos
predict larger values for the ratio $R_b$ and thus diminish the
observed difference  \cite{susy5,susy1,susy3,susy4}.
Complete 1-loop calculations are meanwhile available for
$\Delta r$ \cite{susydelr} and for the $Z$ boson observables
\cite{susy5,susy3,susy4}.

\begin{figure}[htb]
\vspace{-1cm}
\centerline{
\epsfig{figure=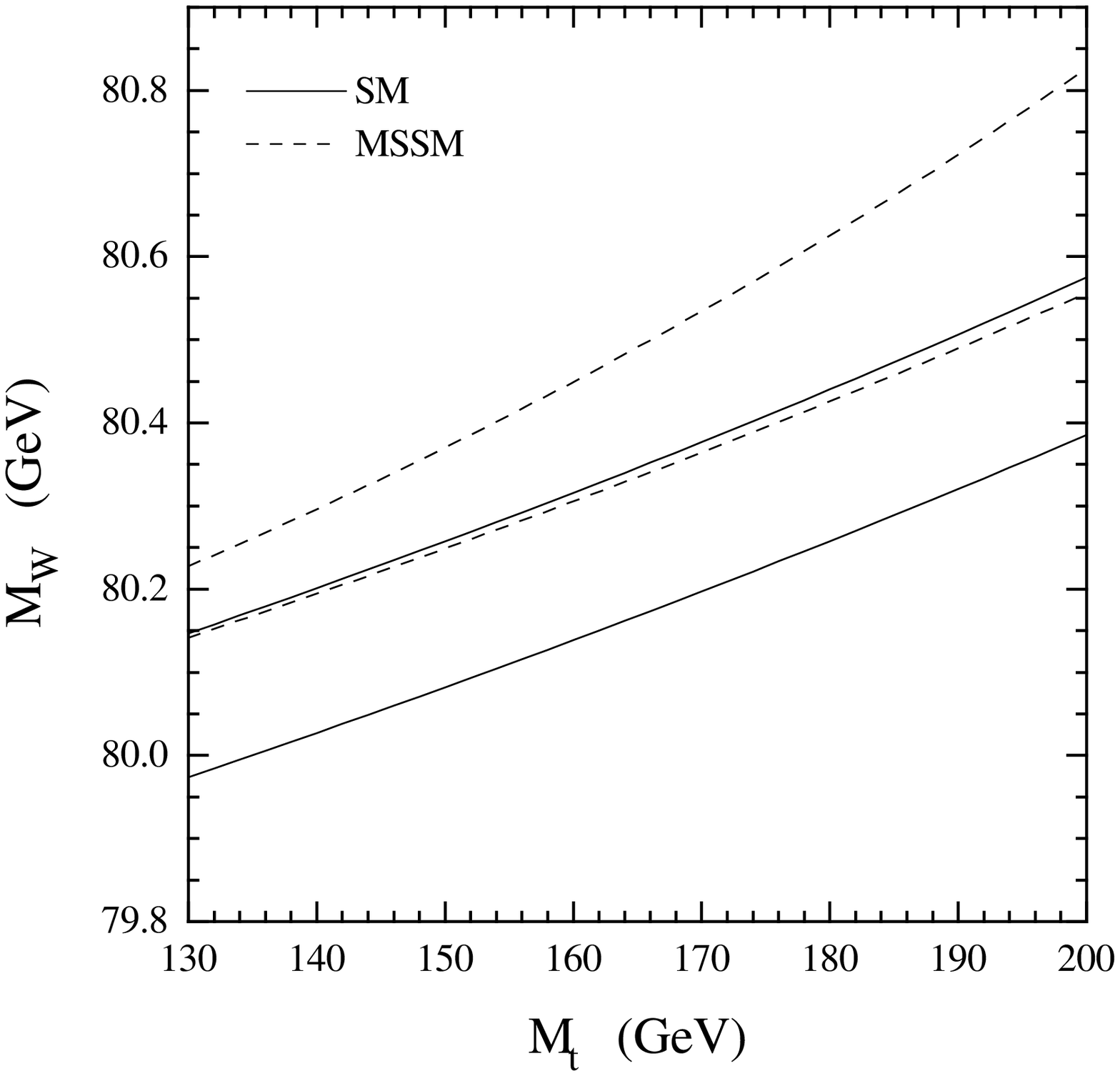,height=15cm,angle=0}}
\vspace{2cm}
\caption{The $W$ mass range in the standard model (-----) and the
         MSSM (- - -). Bounds correspond to the possible situation that 
         no Higgs
         bosons and SUSY particles
         are found at LEP2.} 
\label{susymw}
\end{figure}

Figure \ref{susymw}
displays the range of predictions for $M_W$ in the minimal model
and in the MSSM. Thereby it is assumed that no direct discovery has been
at LEP2. As one can see, precise determinations of $M_W$ and $m_t$
can become decisive for the separation between the  models.

\begin{figure}[htb]
\vspace{-1cm}
\centerline{
\epsfig{figure=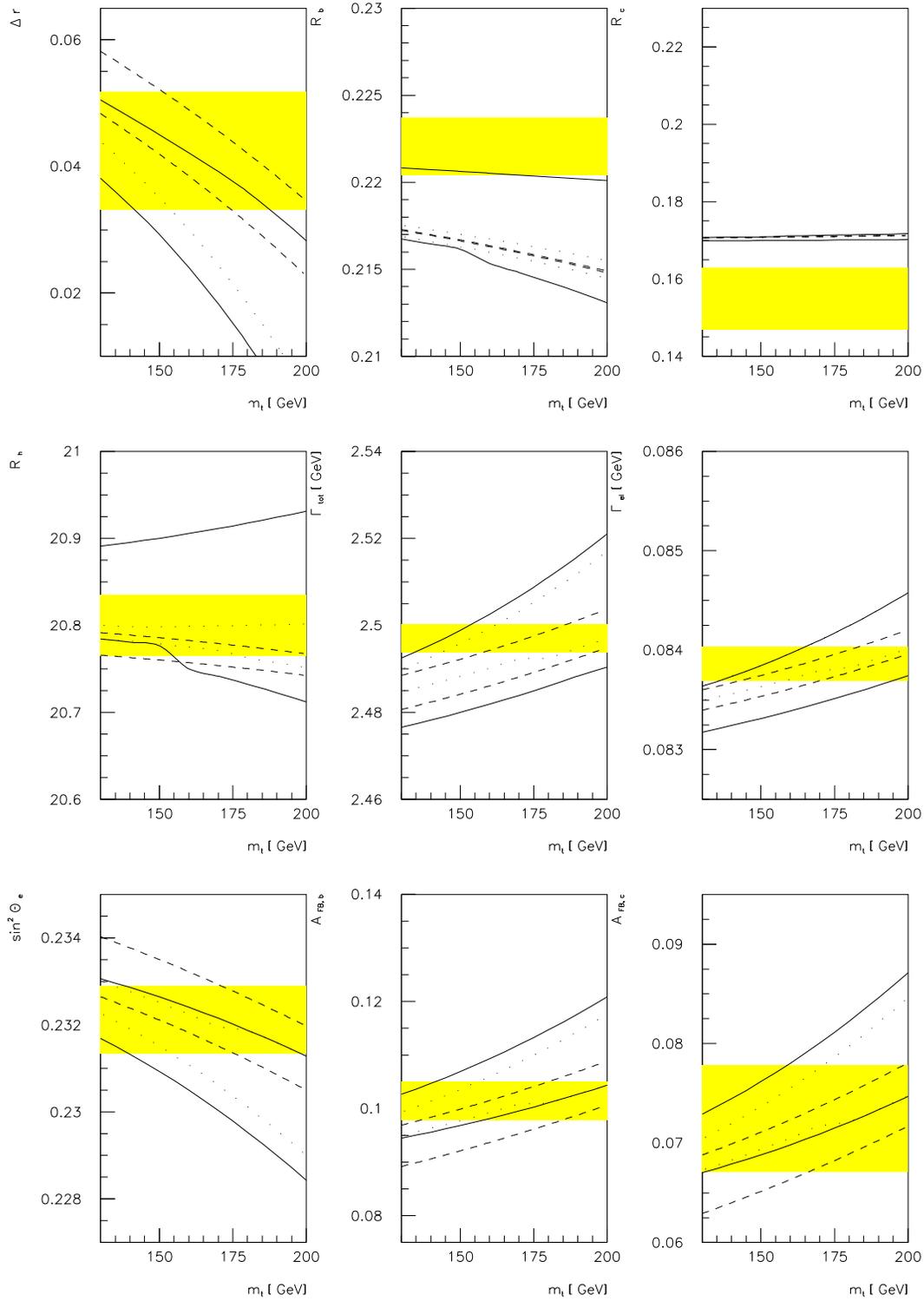,height=20cm,angle=0}}
\caption{Range of precision observables in the standard model (- - -)
         and in the MSSM (---), and present experimental data (dark area).
          The MSSM parameters are restricted by the mass bounds from direct
          searches at LEP I and Tevatron, the dotted lines indicate the 
          bounds to be expected from LEP II.}
\label{zfig5}
\end{figure}

\begin{figure}[htb]
\vspace{-1cm}
\centerline{
\epsfig{figure=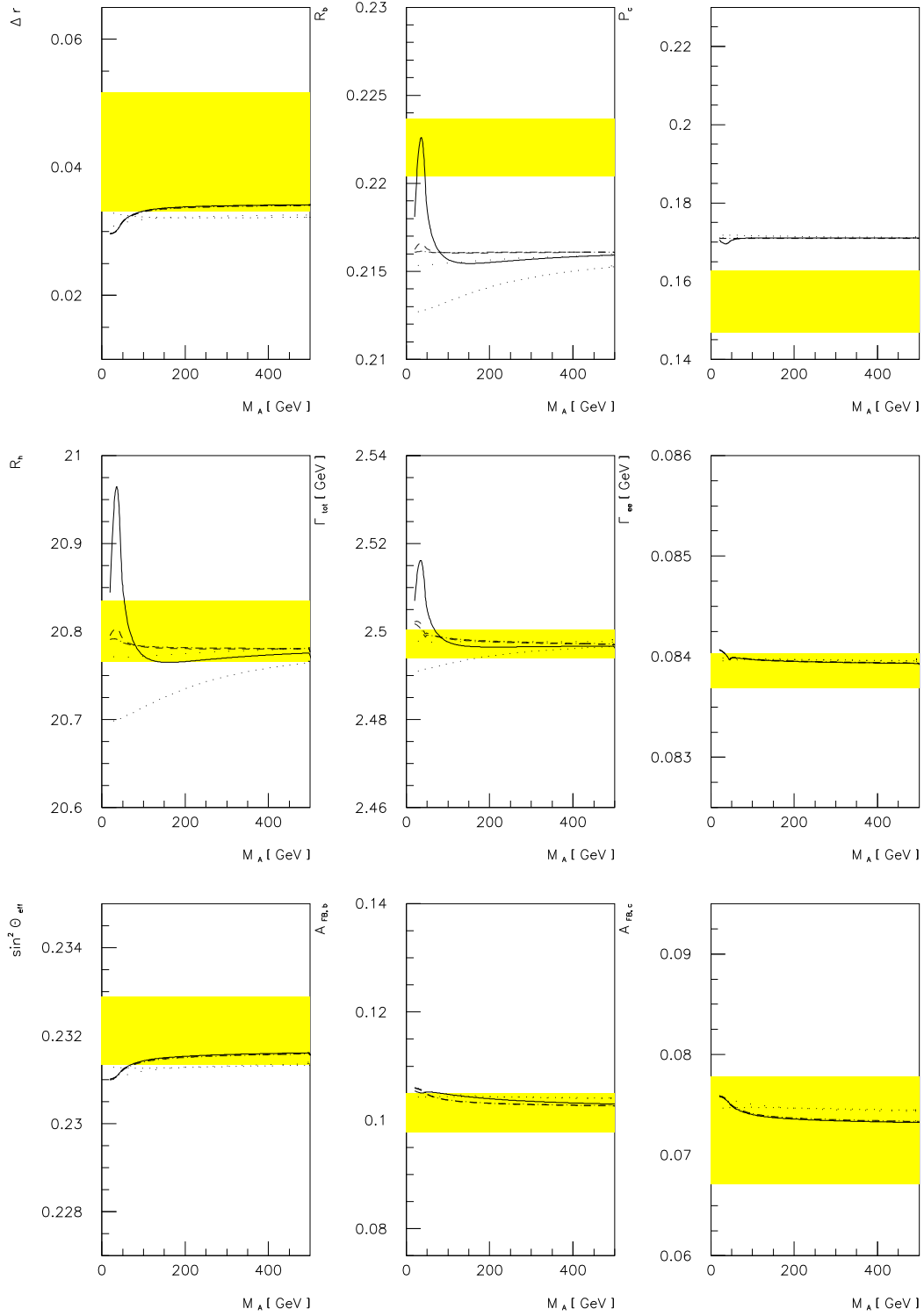,height=20cm,angle=0}}
\caption{Precision observables as function of the pseudoscalar
        Higgs mass $M_A$ for $\tan\beta = 0.7 (\cdot \cdot \cdot), \,
        1.5 (\cdot\;\;\cdot\;\;\cdot), \, 8 (- \cdot -\cdot -), \,
        20$(- - -),  70 (-----). $m_t=174$ GeV, $\al_s=0.123$. 
        $m_{\tilde{l}}=800$ GeV, $m_{\tilde{q}}=500$ GeV,
        $\mu=100$ GeV, $M_2= 300$ GeV.} 
\label{zfig1}
\end{figure}

The range of predictions for $\Dr$ and the $Z$ boson observables in 
the MSSM is visualized in Figure \ref{zfig5}
 (between the solid lines)
 together with the standard model predictions (between the dashed lines)
and with the present experimental data (dark area).
$\tan\beta$ is thereby varied between 1 and 70, the other parameters are
restricted according to the mass bounds from the direct search for
non-standard particles at LEP I and the Tevatron.
From a superficial inspection, one might get the impression that the
MSSM, due to its extended set of parameters, is more flexible to
accomodate also the critical observable $R_b$. A more detailed analysis
shows, however, that those parameter values yielding a ``good'' $R_b$
are incompatible with other data points. An example is given in
Figure \ref{zfig1}: a light $A$ boson together with a large $\tan\beta$
can cure $R_b$, but violates the other hadronic quantities and the
effective leptonic mixing angle. Whereas the hadronic quantities can be
repaired (at least partially) by lowering the value of $\al_s$, the 
mixing angle and $A_{FB}^b$ remain off for small Higgs masses.
Thus, even in the MSSM it is not possible to simultaneously find
agreement with all the individual  precision data.

\smallskip
\noi
The main results can be summarized as follows: \hfill
\\
$\bullet$ $R_c$ can hardly be moved towards the measured range.
 \hfill \\ 
$\bullet$ $R_b$ can come closer to the measured value, in particular
for light $\tilde{t}_R$ and light charginos. \hfill \\
 $\bullet$  $\al_s$ turns out to be smaller than in the 
 minimal model because of  the
reasons explained in the beginning of this section.
\hfill \\ 
$\bullet$ There are strong constraints from the other precision
observables which forbid parameter configurations shifting $R_b$
into the observed $1\sigma$ range.

\begin{table*}[b]
            \bc
 \btab{| l | c | c | c | c | c | }
\hline
\hline
$ \tan\beta $  & $\chi^2$ & $M_{\chi^+_1}/M_{\chi_2^+}$ (GeV)  & $\al_s$ 
            & $m_t$ & $ M_{h^0}$ (GeV)  \\
       &    & & & &   \\
\hline
 1 & 14.7 & 88/93 & 0.110 & 175 &  105  \\
1.2 & 15.6 & 83/99 & 0.112 & 176 & 107 \\
1.6 & 16.8 & 72/110 & 0.115 & 176 & 114 \\
50 & 17.3 & 64/299 & 0.114 & 165 & 50 \\  
\hline
\etab
\ec
\vspace{0.5cm} 
            \caption{Variables for the best fit results
             } \vspace{0.5cm}
\end{table*}

\clearpage

\clearpage

\clearpage

\begin{figure}[hbt]
\vspace{1cm}
\centerline{
\mbox{\epsfxsize8.0cm\epsffile{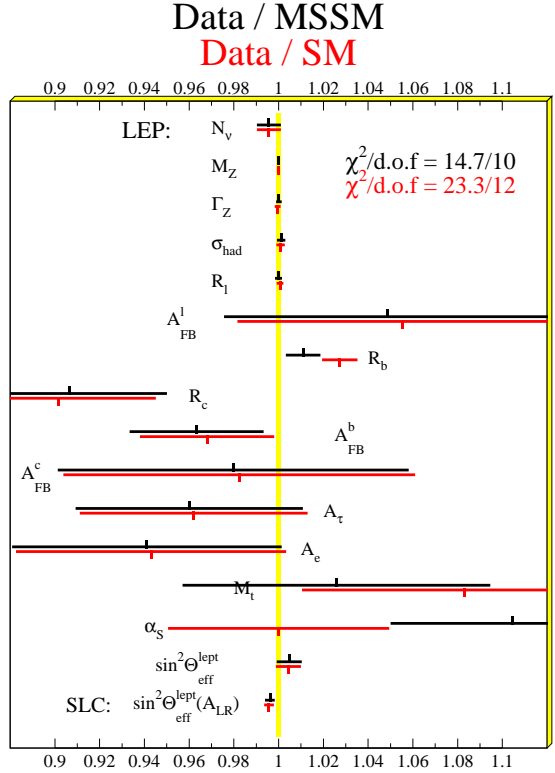}}} 
\caption{Experimental data normalized to the best fit results in
         the SM and MSSM. }
\label{mssm}
\end{figure}

For obtaining the optimized SUSY parameter set, therefore, a global
fit to all the electroweak precision data (including the top
mass measurements)
 has to be performed,
as done in refs.\ \cite{susy4,deboer,brux}. As an example, 
Figure \ref{mssm} displays the
experimental data normalized to the best fit results  
in the SM and MSSM (for $\tan\beta=1$), with 
the data from the 1995 summer conferences \cite{deboer}.
For the SM, $\al_s$ identified with the experimental
number, therefore the corresponding result in Figure 6 is
centered at 1. The most relevant conclusions are: \hfill \\
(i) The difference between the experimental and theoretical value
of $R_b$ is diminished by a factor $\simeq 1/2$, \hfill \\
(ii) the central value for the strong coupling is
 close to the value obtained from deep inelastic
scattering, \hfill \\
(iii) the other observables are practically unchanged, \hfill \\
(iv) the $\chi^2$ of the fit is slightly better than in the minimal
model.

\setlength{\unitlength}{0.7mm}

In table 2 we put together the variables for the best fits (minimum
$\chi^2$) for a few values of low $\tan\beta$ and 
for $\tan\beta=50$. The mass of the nearly righthanded scalar top
is taken at 48 GeV, the other sfermions and Higgs bosons are heavy.
 Only for the large
$\tan\beta$ scenario, we have also a pair of light Higgs bosons
$M_{h^0} \simeq M_{A^0} = 50$ GeV.
The mixing in the scalar top sector is small, but not zero.
A non-diagonal $\tilde{t}$ mass matrix is required to make the $h^0$
sufficiently massive.

In the fits, the SUSY mass parameters $\mu$ and $M_2$ are varied
independently.
In the low $\tan\beta$ regime, the values in table 2 correspond to the
situation $|\mu| \simeq M_2$. The charginos in this case do not have
large mass splittings. They appear as a mixture of wino and Higgsino.
In all cases, the masses for the charginos are not yet excluded 
by the searches at LEP 1.5 \cite{lep2}.

\section{Conclusions}
The experimental data for testing the electroweak theory have
achieved an impressive accuracy.
The observed deviations of several $\sigma$'s in $R_b,R_c,\alr$
reduce the quality of the \sm fits significantly, but the
indirect  determination of $m_t$ is remarkably stable.
Still impressive is the perfect agreement between
theory and experiment for the whole set of the other
precision observables. Supersymmetry can improve the situtation due to an 
enhancement of $R_b$ by new particles in the range below 100
GeV, but it is not possible to accomodate $R_c$.
Within the MSSM analysis, the value for $\al_s$ is close to the one
from deep-inelastic scattering.

\bc
 $ * * * $
\ec
The MSSM fits to the precision data were done
 in a collaboration with A. Dabelstein,
W. de Boer, W. M\"osle, U. Schwickerath \cite{deboer}.
The results were presented partially at the EPS Conference on
High Energy Physics, Brussels 1995 \cite{brux}.  

\bigskip \noi
%

\end{document}